\documentclass[aps,floats,showpacs,twocolumn]{revtex4}
\usepackage{tabularx,graphicx}
\usepackage{epsfig}
\usepackage{amsmath}
\usepackage{bbm}
\usepackage{graphicx}

\begin{document}
%\draft
%\flushbottom
%\twocolumn[
%\hsize\textwidth\columnwidth\hsize\csname @twocolumnfalse\endcsname

\title{Crossover from Luttinger liquid to Coulomb blockade regime
in carbon nanotubes}
\author{S. Bellucci$^a$, J. Gonz\'{a}lez$^b$ and P. Onorato$^{a,c}$ \\}
\address{
        $^a$INFN, Laboratori Nazionali di Frascati,
        P.O. Box 13, 00044 Frascati, Italy \\
        $^b$Instituto de Estructura de la Materia,
        Consejo Superior de Investigaciones Cient\'{\i}ficas, Serrano 123,
        28006 Madrid, Spain\\
        $^c$Dipartimento di Scienze Fisiche,
        Universit\`{a} di Roma Tre, Via della Vasca Navale 84,
        00146 Roma, Italy}

\date{\today}

\begin{abstract}
%\widetext

We develop a theoretical approach to the low-energy properties
of 1D electron systems aimed to encompass the mixed features of
Luttinger liquid and Coulomb blockade behavior observed in
the crossover between the two regimes. For this aim we extend
the Luttinger liquid description by incorporating the effects of
a discrete single-particle spectrum. The intermediate regime is
characterized by a power-law behavior of the conductance, but with
an exponent oscillating with the gate voltage, in
agreement with recent experimental observations. Our construction
also accounts naturally for the existence of a crossover
in the zero-bias conductance, mediating between two temperature
ranges where the power-law behavior is preserved but with
different exponent.

\end{abstract}

%]
\maketitle

%\narrowtext
%\tightenlines

%\newpage

The recent progress in nanotechnology has allowed for a
detailed study of the transport properties of one-dimensional
(1D) electron systems. The discovery of carbon nanotubes (CNs)
in 1991\cite{1}, as a by-product of fullerene production, has
opened a new field of research in mesoscopic physics,
especially because of their potential application to
nanoelectronic devices\cite{dek}.

CN are rolled up sheets of graphite
forming tubes that are only nanometers in diameter, with metallic
or semiconducting properties depending on the roll-up direction
of the sheet. Regarding the metallic CNs, their transport
properties are dictated by the low dimensionality of the system,
which gives rise to strong electronic correlations. Thus, the
effects of the Coulomb interaction become very important in
CNs, and the way they manifest themselves depends mainly on the size of the
system, the temperature, and the quality of the contacts used in
experiments.

When the contacts are not highly transparent, the measurements of the
conductance and the differential conductivity reflect the strong
Coulomb repulsion in CNs. For temperatures that are
typically below 1 K, the zero-bias conductance shows oscillations
characteristic of the so-called Coulomb blockade regime\cite{cb}.
The gate voltage between two peaks is related to the
energy required to overcome the Coulomb repulsion when adding
an electron between the barriers created by the contacts.
In general, the intensity plots in terms of the bias and gate
voltages show in that regime a sequence of diamonds where the
electron tunneling is suppressed, with the size of the diamonds
giving a measure of the energy needed to add an electron to the
system\cite{diam}.

For higher values of the temperature, such that the thermal
energy is much larger than the level spacing, the transport
measurements reflect instead the many-body properties of the
system. The conductance shows then a different kind of
suppression, which arises from the strongly correlated
character of the electron liquid\cite{corr}.
In a 1D system, the electron
quasiparticles are unstable under the slightest interaction
and the low-energy excitations take the form of charge or spin
density fluctuations. In the so-called Luttinger liquid regime,
the absence of electron quasiparticles gives rise to a power-law
suppression of the tunneling density of states (TDOS) near the Fermi
level, $n( \varepsilon ) \sim \varepsilon^{\alpha }$, which
translates in turn into a similar behavior of quantities like
the zero-bias conductance as a function of the temperature,
$G(T) \sim T^{\alpha }$.

Such kind of signatures of Luttinger liquid behavior have been
observed in single-walled\cite{sw} as well as in multi-walled
nanotubes (MWNTs)\cite{mw}.
It has been shown that the zero-bias conductance and the
differential conductivity follow closely a power-law behavior
as functions of the temperature and the bias voltage,
respectively. In the case of the zero-bias conductance, such
power-law dependence holds over a wide range of temperatures,
down to values where the discrete character of the spectrum
starts to become manifest. It is observed then the transition
to the Coulomb blockade regime, where the suppression
of the conductance arises from the single-electron tunneling
mentioned above.

At this point, it would be desirable to have a theoretical
description of the crossover as the temperature decreases from
the Luttinger liquid regime to the Coulomb blockade regime.
There have been experiments like that reported in Ref.
\onlinecite{kanda} in which the intermediate regime has been
explored, measuring the zero-bias conductance at temperatures
where the thermal energy becomes comparable to the level
spacing in the discrete spectrum. The results reported in
that reference provide evidence for the power-law behavior
of the TDOS in MWNTs,
but with an unusual dependence of the exponent $\alpha $ on
the gate voltage. Clear oscillations have been observed in
that plot, with a characteristic scale of the order of 1 V.
Another remarkable observation has been the existence of an
inflection point in the log-log plot of the conductance versus
temperature, for gate voltages corresponding to peak values in
the $\alpha $ exponent\cite{kanda}.

The aim of the present paper is to provide a framework
that may encompass the mixed features of Luttinger liquid and
Coulomb blockade behavior observed in the crossover between
the two regimes. For that purpose we will rely on the 1D
many-body approach\cite{1d}, incorporating at the same time
the discrete character of the spectrum which arises from
the finite size of the system. Thus we will see that
the polarizability, which is proportional to the
single-particle density of states, is modulated at sufficiently
low temperature due to the peaks building the density of states
in the discrete spectrum. Such a modulation is translated to the
scaling behavior of the quasiparticle weight, giving rise to
oscillations in the exponent of the TDOS
as the Fermi level of the system is shifted. We will show that
the period estimated for the oscillations is in agreement with
the experimental observations, and that the proposed framework
is able to account for the features seen in the measurements of the
zero-bias conductance reported in Ref. \onlinecite{kanda}.

In our approach, a central role in distinguishing the
Coulomb blockade from the Luttinger liquid regime is played
by the energy distribution function $n (\varepsilon ;
\varepsilon_F)$. At low temperatures, that function has peaks
that reflect the level spacing $\Delta \varepsilon $ arising
from the level quantization in a finite nanotube. We can
model each peak with a normalized function $f(\varepsilon, n)$,
with a shape depending on the temperature $T$
\begin{equation}
f (\varepsilon, n) = \frac{1}{k_B T \cosh^2
  \left( \frac{\varepsilon - n \Delta \varepsilon }
        { k_B T}     \right)}.
\end{equation}
The energy distribution function is cutoff by the Fermi
energy $\varepsilon_F $, as it is given by
\begin{equation}
n (\varepsilon ; \varepsilon_F ) =
\sum_{n=0}^{\varepsilon_F/\Delta \varepsilon}
  f (\varepsilon, n ).
\label{edf}
\end{equation}

\begin{figure}
\begin{center}
\mbox{\epsfxsize 7.5cm \epsfbox{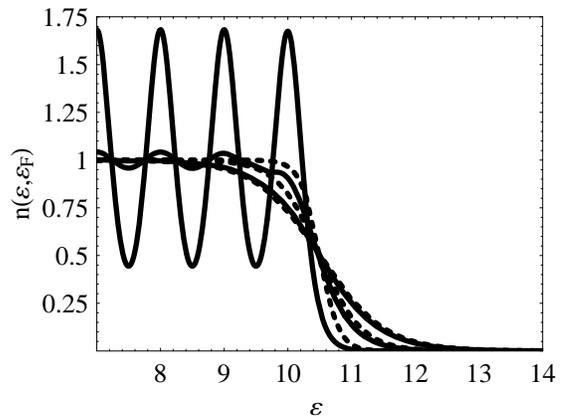}}
\end{center}
\caption{Plot of the energy distribution function $n (\varepsilon
; \varepsilon_F )$ for different values of the temperature
corresponding to $k_B T/\Delta \varepsilon = 0.25, 0.6, 1.0 $.
The solid lines have been obtained from Eq. (\ref{edf}) and the
dashed lines by using the Fermi-Dirac distribution (\ref{cont}).}
\label{one}
\end{figure}

At sufficiently high temperatures, we can neglect the level
spacing and calculate the energy distribution function by
integrating
\begin{eqnarray}
n (\varepsilon ; \varepsilon_F )
 & \approx &
    \int_{-\infty}^{\varepsilon_F/\Delta \varepsilon - 1/2}
     dy  f(\varepsilon, y )                 \nonumber \\
  & = & \frac{1}{1+e^{
   ( \varepsilon -\varepsilon_F +\Delta \varepsilon/2 )/k_B T }}.
\label{cont}
\end{eqnarray}
The last formula gives the usual Fermi-Dirac distribution. We
realize however that the energy distribution function has
in general a more structured shape, that depends on the ratio
between the level spacing $\Delta \varepsilon $ and the thermal
energy $k_B T$ as illustrated in Fig. \ref{one}.

The structure of peaks in the energy distribution function
affects several objects in the many-body theory. This
is the case of the one-loop polarization $\Pi^{(0)} (q, \omega_q)$,
which is given in terms of the electron propagator
$G^{(0)} (k, \omega_k)$ by
\begin{equation}
\Pi^{(0)} (q, \omega_q) = \int \frac{dk}{2\pi }
     \int \frac{d\omega_k }{2\pi } \:
  G^{(0)} (k+q, \omega_k + \omega_q ) G^{(0)} (k, \omega_k).
\end{equation}
This object counts the number of particle-hole excitations that
can be built with momentum $q$ across the Fermi level. Thus, in
the case of a model with linear dispersion
$\varepsilon (k) \approx v_F k$, it becomes proportional
at zero temperature to
\begin{equation}
\int_{-\infty}^\infty dk \: \theta(k_F-k) \theta(q+k-k_F)  = q.
\end{equation}
However, in the model with a distribution function given by
(\ref{edf}), the peaks give rise to a factor in the
integration that reflects the discrete structure
\begin{equation}
  \int_{-\infty}^\infty dk \; n (v_F k; \varepsilon_F )
\left(1 - n (v_F (q+k); \varepsilon_F ) \right) =
q N_T (\varepsilon_F ).
\label{fact}
\end{equation}

The usual form of the 1D polarization is modified then
by the presence of the last factor in (\ref{fact}):
\begin{equation}
\Pi^{(0)} (q, \omega_q) =  \frac{2}{\pi }
  N_s N_T(\varepsilon_F )
 \frac{v_F q^2}{\omega_q^2 - v_F^2 q^2 }
\label{pol}
\end{equation}
with the factor $N_s$ standing for the number of subbands
crossing the Fermi level.
The factor $N_T (\varepsilon_F )$ is itself a periodic
function of $\varepsilon_F $, since the position of the Fermi
energy fixes the portion of the last peak that has to be
integrated in the spectrum, in order to obtain the polarizability.

We can incorporate the polarization (\ref{pol}) in the RPA
evaluation of the electron self-energy
$\Sigma (k, \omega_k)$. This can be expressed as
\begin{eqnarray}
\Sigma (k, \omega_k) & = & i
      \int_{-\Lambda /v_F}^{\Lambda /v_F}
        \frac{dq}{2\pi } \int_{-\infty}^\infty
        \frac{d \omega_q }{2\pi }
      G^{(0)} (k-q, \omega_k - \omega_q)     \nonumber  \\
   &   &  \times \frac{V(q)}{1 - V(q) \Pi^{(0)} (q, \omega_q) },
\end{eqnarray}
where $V(q)$ stands for the 1D Coulomb potential in momentum
space. The high-energy cutoff $\Lambda $ comes from the existence
of a microscopic short-distance scale in the electron system and,
for our purposes, it provides a way of having under control the
scale dependence of the different observables\cite{1d}.

The RPA scheme allows for an accurate description of the behavior
of the quasiparticle weight at low energies, since it gives the
exact result for the anomalous dimension of the electron field
in the 1D system\cite{noi}.
Thus, the self-energy turns out to have terms linear in
$\omega_k$ that depend logarithmically on the cutoff
$\Lambda $. This is the signal of the energy dependence of
the scale $Z^{1/2}$ of the electron wavefunction. The
dressed electron propagator $G(k, \omega_k )$ is given by
\begin{eqnarray}
\frac{1}{G} & = & \frac{1}{G^{(0)}} - \Sigma (k, \omega_k )
                                          \nonumber      \\
 &  \approx  &  Z^{-1}(\Lambda ) (\omega_k - v_F \sigma_x k)
                                         \nonumber       \\
  &   &  + Z^{-1}(\Lambda )
    ( \omega_k - v_F \sigma_x k ) \gamma (g)  \log (\Lambda )
\end{eqnarray}
where $g$ is the effective coupling constant, given in terms
of a suitable average value of the Coulomb potential $\tilde{V}$
by $g = 2 N_s N_T(\varepsilon_F ) \tilde{V} /\pi v_F$. The function
$\gamma (g)$ bears also an explicit dependence on the oscillating
factor $N_T(\varepsilon_F )$, as it is given by
\begin{equation}
\gamma (g) =  \frac{1}{4 N_s N_T(\varepsilon_F ) }
 \left( \sqrt{1+g} + \frac{1}{\sqrt{1+g}} - 2 \right)
\end{equation}

The requirement of cutoff-independence of the renormalized
Green function $G$ is what forces the dependence of $Z $
on the energy scale $\Lambda $, as it is bound then to satisfy
the flow equation
\begin{equation}
\Lambda \frac{d}{d \Lambda} \log Z (\Lambda )  =  \gamma (g).
\label{flow}
\end{equation}
The quantity at the right-hand-side of (\ref{flow}) is actually
the exponent $\alpha $ of the power-law behavior characterizing
the suppression of the quasiparticle weight in the low-energy
limit $\Lambda \rightarrow 0$. The novelty of our framework is
that such an exponent depends now on the Fermi
energy of the system, as the factor $N_T(\varepsilon_F) $ is
sensitive to the position of the Fermi level in the structure
of single-particle levels.

We now ascertain that our approach may account for
the experimental features reported in Ref. \onlinecite{kanda}.
For this purpose, we have first to check that the Fermi energy
shift $\Delta \varepsilon $ producing an oscillation in the
$\alpha $ exponent is consistent with the variation of the gate
voltage applied in the experiment. The dependence of $\alpha $
on the gate voltage $V_g$ has been plotted in Figs. 2(c) and
2(d) of the mentioned reference. There it can be seen that,
in a range of 20 V for the gate voltage, there are about eight
oscillations in the $\alpha $ exponent. The corresponding
change in the Fermi energy can be inferred from the capacitive
coupling between the gate and the MWNT, that
has been estimated to be about $1 \times 10^3$ \cite{kapl}.
Then, each oscillation in the plot should correspond to a
variation in the Fermi energy of about 2 meV. This matches well
the value of $\Delta \varepsilon $
estimated by taking the level spacing from the
finite nanotube length $L$ as a lower bound,
$\Delta \varepsilon \sim hv_F/L \sim 1 \; {\rm meV}$, hence providing
a strong indication that the quantization of the nanotube levels
is at the origin of the oscillations observed in the
$\alpha $ exponent.

\begin{figure}
\begin{center}
\mbox{\epsfxsize 7.5cm \epsfbox{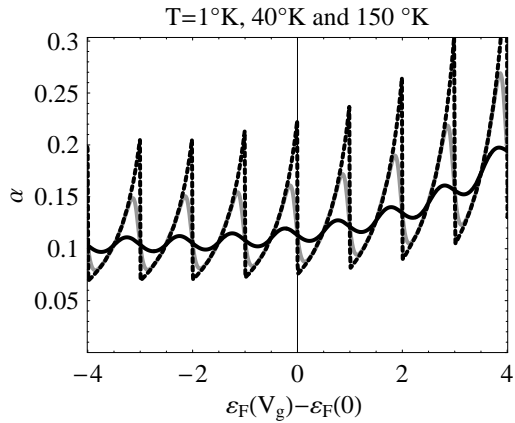}}
\end{center}
\caption{Plot of the exponent $\alpha $ as a function of the Fermi
energy $\varepsilon_F $ (in units of $\Delta \varepsilon$) for
different values of the temperature $T = 1, 40, 150 \; {\rm K}$.}
\label{two}
\end{figure}

The power-law behavior obtained from (\ref{flow}) applies
directly to the TDOS $n (\varepsilon )$,
since we have
\begin{equation}
 n (\varepsilon ) \sim N_s Z (\varepsilon ) \sim
   \varepsilon^{\gamma (g) }.
\end{equation}
We can compare the dependence of the exponent $\gamma (g)$ on
$\varepsilon_F $ with that of the power-law behavior of the
conductance measured in Ref. \onlinecite{kanda}. In our framework,
the exponent depends on the position of the Fermi level due to the
modulation introduced by $N_T(\varepsilon_F )$ in $\gamma (g)$,
but also because of the possible variation of the Fermi velocity
$v_F$ as the Fermi level is shifted. In this respect, we recall
that MWNTs are usually hole-doped, with
several subbands crossing the Fermi level\cite{kr}, so that a
significant decrease in the Fermi velocity at some of the Fermi
points is likely as the top of the lowest subband is approached.
Actually, a drift in the average value of the exponent $\alpha $
is appreciable in Figs. 2(c) and 2(d) of Ref. \onlinecite{kanda},
as the gate voltage runs from $-10$ V to 10 V. We have
incorporated this effect by taking into account a suitable shift
of the Fermi velocity in our model, from $ v_F$ to $5 v_F/6$, as
$\varepsilon_F$ runs over 10 oscillations in the exponent.
Finally, we have taken for the interaction strength $\tilde{V}
\approx 5% (1.5*\pi/2=2.37 se considero la formula a d=1 ottengo \tilde{V}=8*\pi/2)
 v_F$, bearing in mind the screening from the metallic
shells of the MWNT\cite{eg}. The corresponding
plot for $\alpha $ as a function of $\varepsilon_F$ is given
in Fig. \ref{two}. We observe the similarity of the plot with
the experimental curves of Ref. \onlinecite{kanda} for the exponent over the 20 V
range in the gate voltage.

Our approach provides also information about the dependence of
the exponent $\alpha $ on the temperature. It can be seen in Fig.
\ref{two} that the oscillations are significant for $k_B T$
lower than a threshold of the order of $\Delta \varepsilon $.
For higher temperatures, the thermal fluctuations erase the
structure of superposed peaks in (\ref{edf}), what leads in
turn to the disappearance of the oscillations in the $\alpha $
exponent. This agrees with the
experimental observations, which show the existence of an
inflection point $T^{*}$ in the log-log plots of conductance
versus temperature, for values of $V_g$ corresponding to peaks
in the plot of $\alpha $. As reported in Ref. \onlinecite{kanda},
the conductance keeps following a power-law behavior above the
temperature $T^{*}$, but with an exponent significantly
reduced with respect to the value at the peak.

%We find a natural explanation of
%that behavior in our model, since it is already observed in
%Fig. \ref{two} that the peaks of $\alpha $ are most
%significantly affected by the decrease of temperature.

\begin{figure}
\begin{center}
\mbox{\epsfxsize 4.0cm \epsfbox{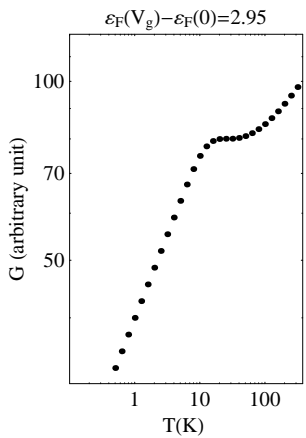}}
\end{center}
\caption{Log-log plot of the conductance as a function of the
temperature.}
\label{three}
\end{figure}

We have checked that, for values of $\varepsilon_F$
corresponding to peaks in the plot of $\alpha $, an inflection
point actually exists in the log-log plot of the conductance.
This is obtained in our framework from the TDOS, by trading the
energy variable in $n(\varepsilon )$ for the
corresponding thermal energy $k_B T$. The resulting dependence
on temperature has been represented in Fig. \ref{three}. The value
of the crossover temperature corresponds to a thermal energy of
the order of $\Delta \varepsilon $, which is consistent with
the value of $T^{*} \approx 30 \; {\rm K}$ that has been found
in the experiment.

We have thus achieved a description that accounts for the
experimental features in the zero-bias conductance reported
in Ref. \onlinecite{kanda}. The discussion carried out in that
paper highlighted the difficulty of finding a theoretical
framework encompassing both the $\alpha $ oscillations and the
existence of the crossover temperature in the conductance.
We have shown, however, that it is possible to extend the
Luttinger liquid description of 1D electron systems by
incorporating the effects of the discrete single-particle
spectrum, thus producing a picture that is relevant to the
mentioned experimental instance. We have seen that the actual
size of the period for the $\alpha $ oscillations can be
explained in our approach, in terms of the level
spacing in the quantized spectrum. The estimated
amplitude of the oscillations is also consistent with the
values measured experimentally. Our theoretical
construction accounts quite naturally for the existence of
a crossover in the zero-bias conductance, keeping track
of its power-law behavior both below and above the crossover
temperature.

To conclude, we have developed a picture that applies to an
intermediate regime between that of Coulomb blockade and the
Luttinger liquid behavior. In the case of a MWNT
with a length of a few microns, for instance, the finite size of
the system leads to a spacing between single-particle levels
which is of order $\sim 0.1 \: {\rm meV}$. Thus, for
temperatures which are $\lesssim 1 \: {\rm K}$ we have the
Coulomb blockade regime, with the characteristic oscillations
in the conductance as the gate voltage is varied. For
temperatures well above that value, the conductance shows
a power-law suppression as a function of the temperature,
typical of Luttinger liquid behavior. We have shown that the
intermediate regime can be characterized by a power-law
behavior of the conductance, but with an exponent that
oscillates as a function of the gate voltage, in agreement with
observations reported in Ref. \onlinecite{kanda}. This
reassures the robustness of the Luttinger liquid picture for
the description of the CNs, down to the low-energy
scales where the collective excitations of the electron
system begin to fade away.

\acknowledgements

%{\center \bf Acknowledgements}
 \noindent J. G. acknowledges
the financial support of the Ministerio
de Educaci\'on y Ciencia (Spain) through grant
BFM2003-05317.


\begin{thebibliography}{99}


\bibitem{1}
S. Iijima, Nature {\bf 354}, 56 (1991).

\bibitem{dek}
C. Dekker, Phys. Today {\bf 52}, 22 (1999).

\bibitem{cb}
M. Bockrath, et al., Science {\bf 275}, 1922 (1997).
S. J. Tans, et al., Nature {\bf 386}, 474 (1997).

\bibitem{diam}
H. W. Ch. Postma, et al.,
Science {\bf 293}, 76 (2001).
M. R. Buitelaar, et al., Phys. Rev. Lett. {\bf 88}, 156801 (2002).

\bibitem{corr}
R. Egger and A. O. Gogolin, Phys. Rev. Lett. {\bf 79}, 5082
(1997).
C. Kane, L. Balents and M. P. A. Fisher, Phys. Rev. Lett.
{\bf 79}, 5086 (1997).

\bibitem{sw}
M. Bockrath, et al., Nature {\bf 397}, 598 (1999).
Z. Yao, et al.,
Nature {\bf 402}, 273 (1999).

\bibitem{mw}
C. Sch\"{o}nenberger, et al.,
Appl. Phys. A {\bf 69}, 283 (1999).
A. Bachtold, et al., Phys. Rev. Lett. {\bf 87},
166801 (2001).

\bibitem{kanda}
A. Kanda, et al., Phys. Rev. Lett.
{\bf 92}, 36801 (2004).

\bibitem{1d}
V. J. Emery, in {\it Highly Conducting One-Dimensional Solids},
ed. J. T. Devreese, R. P. Evrard and V. E. Van Doren (Plenum, New
York, 1979).
J. Solyom, Adv. Phys. {\bf 28}, 201 (1979).

\bibitem{noi}
S. Bellucci and J. Gonz\'alez, Eur. Phys. J. B {\bf 18}, 3
(2000). S. Bellucci,
{\it Path Integrals from peV to TeV}, eds. R. Casalbuoni, et al.
(World Scientific, Singapore, 1999) p.363, hep-th/9810181.
S. Bellucci and J. Gonz\'alez, { Phys. Rev. B}  {\bf 64},
201106(R) (2001). S. Bellucci, J. Gonz\'alez and P. Onorato,
{Nucl. Phys. B} {\bf 663} [FS], 605 (2003).
S. Bellucci, J. Gonz\'alez and P. Onorato, Phys. Rev. B {\bf 69},
085404 (2004).

\bibitem{kapl}
A. Kanda, et al., Appl. Phys. Lett. {\bf 79}, 1354 (2001).

\bibitem{kr}
M. Kr\"uger, et al., Appl. Phys. Lett. {\bf 78}, 1291 (2001).

\bibitem{eg}
R. Egger, Phys. Rev. Lett. {\bf 83}, 5547 (1999).





\end{thebibliography}
\end{document}